\documentstyle[preprint,aps]{revtex}
\begin{document}
\tightenlines
\draft
\preprint{
\parbox{4cm}{
\baselineskip=12pt
KEK-TH-770\\ 
May, 2001\\
\hspace*{1cm}
}}
\title{Dynamical Doublet-Triplet Higgs Mass Splitting} 
\author{ Ryuichiro Kitano $^{ab}$ 
  \thanks{e-mail: ryuichiro.kitano@kek.jp} 
     and Nobuchika Okada $^a$ 
  \thanks{e-mail: okadan@camry.kek.jp}}  

\address{$^a$Theory Group, KEK, Oho 1-1, Tsukuba, Ibaraki 305-0801, Japan}
\address{$^b$Department of Particle and Nuclear Physics, 
         The Graduate University for Advanced Studies, 
         Oho 1-1, Tsukuba, Ibaraki 305-0801, Japan}        
%
\maketitle
\vskip 2.5cm
\begin{center}
{\large Abstract}
\vskip 0.5cm
\begin{minipage}[t]{14cm}
\baselineskip=19pt
\hskip4mm
We propose a new mechanism toward the solution 
 to the doublet-triplet Higgs mass splitting problem 
 in the supersymmetric grand unified theory. 
Our model is based on the gauge group $SU(5)_H \times SU(5)_{GUT}$, 
 where $SU(5)_H$ and $SU(5)_{GUT}$ are 
 a new strong gauge interaction and the ordinary grand unified 
 gauge group, respectively. 
The doublet-triplet Higgs mass splitting is realized
 through the quantum deformation of moduli space 
 caused by the strong $SU(5)_H$ gauge dynamics. 
The low energy description of our model is given by 
 the minimal supersymmetric standard model. 
%
\end{minipage}
\end{center}
\newpage
\section{Introduction} 

Supersymmetric extension is one of the the most promising ways 
 in order to provide a solution to the gauge hierarchy problem 
 in the standard model \cite{review}. 
The minimal version of this extension of the standard model is called 
 the minimal supersymmetric standard model (MSSM). 
Interestingly, the experimental data support 
 the unification of three gauge couplings 
 at the scale $M_{GUT} \sim 10^{16} \mbox{GeV}$ 
 with only the MSSM particle contents \cite{unification}. 
This fact seems to strongly suggest the original idea 
 of the grand unified theory (GUT) \cite{gut} 
 together with supersymmetry (SUSY), that is, SUSY GUT. 
At high energy, the theory may be described 
 by a simple gauge group such as $SU(5)$ 
 into which the standard model is embedded. 

However, the GUT suffers from a serious problem, namely, 
 the doublet-triplet Higgs mass splitting problem. 
While the doublet parts of the Higgs superfields  
 in $\overline{\bf 5}$ and ${\bf 5}$ representations 
 (in terminology of $SU(5)$ GUT) 
 should have the electroweak scale mass, 
 the triplet part Higgs masses have to be in the grand unification scale 
 in order to avoid rapid proton decay \cite{proton} 
 and not to destroy the successful gauge coupling unification. 
In a simple treatment, a parameter in the theory should be chosen 
 with accuracy of $10^{-14}$ to realize this mass splitting. 
Although, in a supersymmetric theory, 
 this fine-tuning is ``technically natural''  
 because of the non-renormalization theorems \cite{non}, 
 this is basically the same problem as the original gauge hierarchy problem. 
Thus, the theory may not be compelling 
 as a resolution of the hierarchy problem, 
 unless this mass splitting is naturally realized 
 without the fine-tuning. 
 
There have been many efforts to solve the doublet-triplet Higgs 
 mass splitting problem. 
These are roughly classified into two approaches. 
One is the dynamical approach such as 
 the sliding singlet mechanism \cite{slide}. 
The other is the group theoretical one such as, for example, 
 the missing partner mechanism \cite{missingI} \cite{missingII}, 
 the Dimopoulos-Wilczek mechanism \cite{DW}, 
 the idea of Higgs doublets as pseudo Nambu-Goldstone bosons \cite{NG}. 
More recently, an alternative approach was investigated 
 based on the extra-dimensional theories \cite{extra}. 

In this paper, we propose a new mechanism which realizes 
 the doublet-triplet Higgs mass splitting 
 through the strong gauge dynamics. 
Our model is based on the gauge group $SU(5)_H \times SU(5)_{GUT}$, 
 where $SU(5)_H$ is the strong gauge interaction 
 additionally introduced to the ordinary GUT gauge group $SU(5)_{GUT}$. 
At low energy, the $SU(5)_H$ gauge interaction becomes strong, 
 and the theory is described by the effective field $M$ 
 of the {\bf{25}} (={\bf{1}}+{\bf{24}}) representation 
 under the $SU(5)_{GUT}$ gauge group.  
In our model, this effective field has the effective Yukawa coupling 
 among the pair of Higgs superfields ($\bar{H}$ and $H$) 
 such as $W=\bar{H} M H$. 
 This is the unique source for the Higgs masses. 
The strong $SU(5)_H$ dynamics causes the quantum deformation of moduli space, 
 whose SUSY vacuum condition requires 
 that the effective field should have, at least, two zero eigenvalues. 
On the desired vacuum where the GUT gauge symmetry is correctly broken 
 to the MSSM gauge group, 
 this means that the strong dynamics requires 
 that the pair of Higgs doublets are massless, 
 but the pair of Higgs triplets are heavy 
 through the effective Yukawa coupling. 
Therefore, the doublet-triplet Higgs mass splitting is realized 
 through the effect of the strong gauge dynamics. 
This is a new mechanism, we propose, 
 to solve the doublet-triplet Higgs mass splitting problem. 
 
In the next section, we discuss the dynamical sector of our model 
 and the realization of the doublet-triplet Higgs mass splitting 
 through the strong gauge dynamics. 
Our model has some non-renormalizable superpotentials 
 which are necessary for our mechanism to work. 
The origins of them are discussed in Sec.~III. 
Also, a concrete model as an example is presented, 
 in which the $SU(5)_{GUT}$ gauge symmetry is correctly broken to 
 the standard model gauge group 
 and the light particle contents are the same as that of the MSSM. 
In Sec.~IV, we summarize our discussion and give some comments. 

\section{dynamical doublet-triplet Higgs mass splitting}
 
In this section, we discuss the dynamical sector of our model 
 in which the doublet-triplet Higgs mass splitting is realized 
 through the strong gauge dynamics. 
This sector is based on the gauge group $SU(5)_H \times SU(5)_{GUT}$ 
 with the particle contents as follows. 
\begin{center}
\begin{tabular}{cccc}
 \hspace{1cm}& $~SU(5)_H$~ & $~SU(5)_{GUT}~$ & $U(1)_R$     \\
$ \bar{Q} $   &  $\overline{\bf{5}}$  & $\overline{\bf{5}}$ & $ 2/5$   \\
$    Q  $   &  $\bf{5}$             & $\bf{5}$   & $2/5$   \\
$ \bar{H} $   &  $ \bf{1}$             & $\overline{\bf{5}}$  & $4/5$   \\   
$      H  $   &  $ \bf{1}$             & $ \bf{5}$    & $2/5$   
\end{tabular}
\end{center}
Here, $\bar{H}$ and $H$ are the ordinary Higgs superfields 
 in the $SU(5)_{GUT}$ theory, and $U(1)_R$ is the R-symmetry.
\footnote{ This R-symmetry is anomalous with respect to the $SU(5)_H$ 
 gauge interaction, and broken down to the discrete symmetry $Z_5$ 
 (see Eq.~(\ref{total})). 
 It is possible to introduce the discrete $Z_5$ symmetry from the beginning,
 instead of the $U(1)_R$ symmetry. 
 In any case, we obtain the same results. }
Note that any renormalizable superpotentials are forbidden   
 by the R-symmetry. 
In the following, we consider some non-renormalizable superpotentials 
 consistent with all the symmetries. 
Their origins are discussed in Sec.~III 
 based on all the renormalizable superpotentials. 

Let us first consider the superpotential 
 of the form 
\begin{eqnarray}
 W = \bar{b} \bar{Q}^5 + b Q^5 \; ,  
 \label{BB}
\end{eqnarray} 
 where $\bar{Q}^5$ and $Q^5$ denote the contraction of the gauge
 indices by the epsilon tensors, 
 and $\bar{b}$ and $b$ are the constants 
 with the mass dimension $-2$.
At low energy, the $SU(5)_H$ gauge interaction becomes strong. 
We can describe the low energy effective theory 
 by the method of Seiberg and co-workers \cite{seiberg}.
The moduli space is deformed by the strong gauge dynamics 
 so as to satisfy the condition,  
 $\det M/\Lambda^3  - \bar{B} B = \Lambda^2$, 
 where $\Lambda$ is the dynamical scale of the gauge interaction 
 whose scale is assumed as $ \Lambda \sim M_{GUT}$, 
 and $M$, $\bar{B}$ and $B$ are the mesonic and baryonic 
 effective fields, which are singlet under $SU(5)_H$, as follows. 
\begin{center}
\begin{tabular}{ccccc}
 \hspace{1cm} &    &    & $~SU(5)_{GUT}~$ & $U(1)_R$    \\
$    M  $     & $ \sim $ & $ (\bar{Q} Q)/ \Lambda  $ &   \bf{1}+\bf{24} 
& $ 4/ 5 $  \\
$ \bar{B} $   & $ \sim $ & $ \bar{Q}^5/ \Lambda^4$ &    \bf{1} &$2$    \\   
$      B  $   & $ \sim $ & $  Q^5/ \Lambda^4  $ &    \bf{1} & $2$ 
\end{tabular}
\end{center}
The dynamically generated constraint is incorporated in the superpotential 
 by introducing the Lagrange multiplier superfield $X$, 
 and the total effective superpotential is given by 
\begin{eqnarray}
 W = \Lambda^4 \bar{b} \bar{B} + \Lambda^4  b B 
   + X \left(\frac{\det M}{\Lambda^3} - \bar{B} B - \Lambda^2 \right) \; . 
 \label{total}
\end{eqnarray}

Considering the SUSY vacuum conditions for the heavy fields, 
 $\bar{B}$, $B$ and $X$, and integrating them out, 
 we obtain a more convenient form of the effective superpotential such that 
\begin{eqnarray}
 W = \pm \Lambda^4 \left(\bar{b} b \right)^{1/2}  
     \left( \frac{\det M}{\Lambda^3}- \Lambda^2 \right)^{1/2}  \; .
 \label{eff}
\end{eqnarray}
The SUSY vacuum condition for $M$ leads to 
\begin{eqnarray}
  \left(\det M \right)  M^{-1} = 0 \; , 
 \label{VAC}
\end{eqnarray}
 which means that the effective field $M$ should have, at least, 
 two zero eigenvalues, namely, 
 $ M  = \mbox{diag}  (a_1, a_2, a_3, 0, 0)$. 
Although, in general, $a_i \; (i=1,2,3)$ are arbitrary constants, 
 they are assumed to be $a_1=a_2=a_3=v \sim M_{GUT}$ 
 in order to obtain the correct $SU(5)_{GUT}$ symmetry breaking pattern, 
 $SU(5)_{GUT} \rightarrow SU(3)_c \times SU(2)_L \times U(1)_Y$.
We will present a model to realize this situation 
 in the next section. 

Now, we show that the doublet-triplet Higgs mass splitting 
 is realized through the effects of the strong gauge dynamics. 
Suppose the tree level superpotential such as
\footnote{ The absence of the term  
 $ \bar{H}_i H^i (\bar{Q}_{\alpha j} Q^{\alpha j}) $ 
 is crucial for our mechanism to work. 
 The reason why we do not introduce the term will be explained 
 in the next section. } 
\begin{eqnarray} 
 W = \lambda   H^i (\bar{Q}_{\alpha i} Q^{\alpha j}) \bar{H}_j \; , 
  \label{HQQH} 
\end{eqnarray}
 where $\alpha$ and $i,j$ are the gauge indices of 
 $SU(5)_H$ and $SU(5)_{GUT}$, respectively, 
 and $\lambda$ is a constant with the mass dimension $-1$. 
At low energy, this superpotential is described in terms 
 of the effective field $M$, and leads to the Yukawa coupling 
\begin{eqnarray} 
    W= \left(\lambda \Lambda \right)\; \bar{H} M H  \; .
  \label{HMH} 
\end{eqnarray}
Note that, as discussed above, 
 the strong gauge dynamics requires the eigenvalues for $M$ 
 such that $M = \mbox{diag} (v, v, v, 0,0)$. 
Therefore, the doublet-triplet Higgs mass splitting 
 is realized through the effect of the strong gauge dynamics. 
This is the point of this paper. 

The effective description of our model is similar to the models 
 with the sliding singlet mechanism \cite{slide}. 
In these models, the same superpotential as Eq.~(\ref{HMH}) is introduced, 
 and $M = \mbox{diag} (v, v, v, 0,0)$ is realized 
 through the SUSY vacuum conditions, 
 $\partial W/ \partial \bar{H}=0$ and $\partial W/ \partial H=0$, 
 providing the vacuum expectation value (VEV)  
 for the doublet part of the Higgs fields.  
However, these models suffer from the tadpole problem 
 \cite{tadpoleI} \cite{tadpoleII}, 
 once the SUSY breaking effects in supergravity scenario 
 are taken into account. 
Note that the direction to the non-zero F-terms, 
 $\partial W/ \partial \bar{H}$ and $\partial W/ \partial H$, 
 is bounded by the VEV of the Higgs doublets (the electroweak scale), 
 and is almost flat compared with the typical SUSY breaking scale. 
Thus, the vacuum destabilization occurs, once the tadpole terms 
 of the singlet field are induced through the supergravity effects. 
As a result, the SUSY vacuum on which the doublet Higgs superfields 
 are massless is smoothed out, 
 and the sliding singlet mechanism cannot work.
\footnote{ There is a possibility to avoid this problem 
 \cite{LSUSY} 
 by considering the low scale SUSY breaking scenario,  
 such as the gauge mediated SUSY breaking scenario 
 \cite{GMSB}.} 
On the other hand, note that, in our model, 
 the massless-ness of the doublet Higgs superfields is required 
 by the strong gauge dynamics whose scale is much larger 
 than the typical SUSY breaking scale. 
Thus, the stability of the SUSY vacuum is rigidly ensured, 
 and our model is free from the tadpole problem. 
This is the crucial difference of our mechanism from 
 the sliding singlet mechanism. 

\section{A concrete model} 

Now we present a concrete model as an example, 
 in which our mechanism discussed in the previous section can work 
 and there is the desired SUSY vacuum. 

Let us first discuss the origins of the non-renormalizable superpotentials 
 in Eqs.~(\ref{BB}) and (\ref{HQQH}). 
In the following discussion, we always consider the case 
 that all the fundamental superpotentials are renormalizable. 
In this sense, the non-renormalizable terms are effective ones 
 which are originated from integrating out heavy superfields 
 according to the SUSY vacuum conditions. 
What types of non-renormalizable superpotentials are induced 
 depends on the particle contents of the model. 
The heavy particle contents, in our model, are as follows. 
\begin{center}
\begin{tabular}{ccccc}
 \hspace{1cm} & $~SU(5)_H~$ & $~SU(5)_{GUT}~$ & $U(1)_R$ &    \\
$ \bar{\Phi}_1 $ & $\overline{\bf{10}}$ & $ \overline{\bf{10}} $ & $ 4/5$  \\
$ \bar{\Phi}_2 $ & $\overline{\bf{10}}$ & $ \overline{\bf{10}} $ & $ 6/5$  \\
$ \Phi_1      $ & $ \bf{10}$ & $ \bf{10} $ & $ 6/5$  \\
$ \Phi_2      $ & $ \bf{10}$ & $ \bf{10} $ & $ 4/5$  \\
$ \bar{A}     $ & $ \overline{\bf 5} $  & $ {\bf 1} $ & $ 4/5 $ \\
$      A      $ & $ {\bf 5}          $  & $ {\bf 1} $ & $ 6/5 $ 
\end{tabular} 
\end{center}

The general renormalizable superpotential including 
 $\bar{\Phi}_j$ and $\Phi_j$ ($j=1,2)$ is found to be 
\begin{eqnarray}
 W &=& \bar{Q} \bar{Q} \Phi_1 + \bar{Q} \bar{\Phi}_1 \bar{\Phi}_1 
     + m_1  \bar{\Phi}_1 \Phi_1  \nonumber \\ 
   &+& Q Q \bar{\Phi}_2 + Q \Phi_2 \Phi_2 
     + m_2  \bar{\Phi}_2 \Phi_2    \; , 
 \label{orgBB} 
\end{eqnarray}
 where the gauge indices are appropriately contracted, 
 $m_1$ and $m_2$ are the mass terms, 
 and we take all the Yukawa coupling constants to be $1$, 
 for simplicity. 
Since $\bar{\Phi}_j$ and $\Phi_j$ are charged under the $SU(5)_H$ gauge group, 
 they should be heavy enough, at least, heavier 
 than the dynamical scale $\Lambda$ 
 in order not to change the $SU(5)_H$ dynamics 
 discussed in the previous section. 
After integrating the heavy superfields out, 
 we obtain the effective superpotential in Eq.~(\ref{BB}) 
 with the relations $\bar{b}=1/m_1^2$ and $b = 1/m_2^2$. 
\footnote{ 
 Even if we take $m_1 \sim m_2 \sim M_P$ ($M_P$ is the Planck mass), 
 the coefficients of the tadpole terms for $\bar{B}$ and $B$ 
 are much larger than the typical SUSY breaking scale. 
 No destabilization occurs for the SUSY vacuum we discussed, 
 even if the SUSY breaking effects are taken into account.}  

The effective superpotential in Eq.~(\ref{HQQH}) is induced as follows. 
The general renormalizable superpotential 
 including $\bar{A}$ and $A$ is given by 
\begin{eqnarray}
 W =  \bar{A} Q \bar{H} +  A \bar{Q} H + m_A \bar{A} A \; , 
\end{eqnarray}
 where $m_A$ is the mass parameter, and 
 we take all the Yukawa couplings to be $1$, for simplicity. 
This mass term should be large enough, at least, larger than $\Lambda$. 
We assume that $m_A$ is slightly larger than $\Lambda$. 
Integrating out $\bar{A}$ and $A$ leads to Eq.~(\ref{HQQH}) 
 with the relation $\lambda=-1/m_A \sim -1/\Lambda $.  
Note that the term $ \bar{H}_i H^i (\bar{Q}_{\alpha j} Q^{\alpha j})$ 
 which destroys our mechanism is not generated 
 with the above particle contents. 
This is the reason why we introduced only the superpotential 
 of Eq.~(\ref{HQQH}) in the previous section.  

Finally, let us consider the superpotential 
including the effective field $M$. 
Through this superpotential, we can obtain the desired SUSY vacuum  
 where our mechanism of the doublet-triplet Higgs mass splitting 
 can work and the correct pattern of the GUT gauge symmetry breaking 
 is realized. 
We propose the following superpotential as an example, 
\begin{eqnarray}
 W = \mbox{tr} \left[ \left(\bar{Q} Q \right) \bar{\Sigma} \right] 
  + \mbox{tr} \left[ \Sigma^2 \bar{\Sigma} \right]
  +  Z \left( 
  \mbox{tr} \left[ \Sigma^2 \right] + \mbox{tr} 
  \left[ \bar{Q} Q \right] - m^2  \right) \; , 
 \label{sigma} 
\end{eqnarray}
 where $\bar{\Sigma}$, $\Sigma$ and $Z$ are superfields 
 having charges $({\bf 1},{\bf{24}},  6/5)$, 
 $({\bf 1},{\bf{24}},  2/5)$ 
 and $({\bf 1},{\bf 1},  6/5)$, respectively, 
 under $SU(5)_H \times SU(5)_{GUT} \times U(1)_R$, 
 all the Yukawa coupling constants are taken to be $1$, for simplicity, 
 and the explicit $U(1)_R$ symmetry breaking parameter $m$  
 is introduced. 
Note that, except the $m$ term, the above superpotential is general one 
 consistent with all the symmetries. 
The first and the fourth terms lead to the mass terms 
 for the effective field $M$ in the effective field description. 
The parameter $m$ is assumed to be the order of the GUT scale. 
Although this superpotential is simple, it should be regarded 
 as the effective one because of the introduction of $m$. 
We give a comment on this point in the last section. 

Considering the superpotentials of Eqs.~(\ref{eff}) and (\ref{sigma}),  
 we find the SUSY vacuum such that 
 $ Z =0$, $\bar{\Sigma}=0$, $M=\mbox{diag} (v,v,v,0,0) $ and 
 $\Sigma= \mbox{diag} 
 (2 \sigma, 2 \sigma, 2 \sigma,  -3 \sigma, -3 \sigma)$ 
 with $v = m^2/ 9 \Lambda \sim M_{GUT}$ 
 and $\sigma =\pm m /3\sqrt{5} \sim M_{GUT}$.
\footnote{There are other SUSY vacua 
 where, for example, the triplet Higgs superfields are massless 
 but the doublet Higgs superfields are heavy. 
 We do not consider such a case since it is out of our interests.} 
Now, we obtain the desired SUSY vacuum  
 where our mechanism of the doublet-triplet Higgs mass splitting can work, 
 the GUT gauge symmetry is correctly broken to the standard model one, 
 and all the light particle contents are the same as that of the MSSM.  

Since, at low energy $ \leq M_{GUT}$, 
 the above model is described by the MSSM, 
 the standard model gauge couplings are successfully unified 
 at $M_{GUT} \sim 10^{16}$ GeV. 
At high energy larger than the GUT scale, 
 the MSSM gives place to the $SU(5)_{GUT}$ GUT. 
Although the particle contents introduced above 
 make the $SU(5)_{GUT}$ GUT asymptotic non-free, 
 the gauge coupling remains in the perturbative regime 
 below the Planck scale with appropriate large mass parameters, 
 for example, $m_1, m_2 \gg M_{GUT}$.

\section{Summary and comments} 

We proposed a new mechanism towards the solution of 
 the doublet-triplet Higgs mass splitting problem 
 in the supersymmetric grand unified theory. 
Our model is based on the gauge group $SU(5)_H \times SU(5)_{GUT}$ 
 with particle contents introduced in the sections II and III. 
All the fundamental superpotentials in our model are 
 renormalizable and general, which are consistent with all the symmetries. 
In the present model, we obtained the desired SUSY vacuum, 
 in which the doublet-triplet Higgs mass splitting is realized 
 through the effect (the quantum deformation of moduli space) 
 by the strong $SU(5)_H$ gauge dynamics, 
 and the correct GUT gauge symmetry breaking takes place. 
At low energy, our model connects 
 with the minimal supersymmetric standard model. 

Here, we give some comments on more detailed features of our model. 
Let us first discuss the origin of the explicit $U(1)_R$ breaking term 
 in Eq.~(\ref{sigma}). 
This term is necessary to realize the correct breaking pattern 
 of the $SU(5)_{GUT}$ and to make 
 the fields except the MSSM particle contents heavy. 
One possibility to provide this term is 
 to introduce another strong gauge dynamics. 
For example, consider a sector based on the gauge group 
 $SU(2)_H$ with four doublet superfields $Q_i\; (i=1,2,3,4)$ 
 having the $U(1)_R$ charge $ 2/5 $. 
In this case, the general renormalizable superpotential at the tree level 
 is of the form 
\begin{eqnarray}
 W = - Z \left[ Q_i Q_j \right]  \; , 
 \label{su2}
\end{eqnarray} 
 where $[\; ]$ denotes the contraction of the $SU(2)_H$ indices 
 by the epsilon tensor. 
The strong $SU(2)_H$ gauge dynamics causes the quantum deformation 
 of moduli space, and the constraint, 
 $\mbox{Pf}\left[ Q_i Q_j \right]= \Lambda_2^4$, comes out \cite{seiberg}. 
Here, $\Lambda_2$ is the dynamical scale of the $SU(2)_H$ gauge interaction. 
Under this constraint, the above superpotential leads to 
 the $U(1)_R$ symmetry breaking term $m = \Lambda_2$ in Eq.~(\ref{sigma}). 
This is the dynamical origin of the $SU(5)_{GUT} \times U(1)_R$  
 symmetry breaking through Eqs.~(\ref{sigma}) and (\ref{su2}). 
It is possible to construct other examples. 
 
In this paper, we have considered only the SUSY vacuum. 
If the SUSY breaking effects are taken into account, 
 the soft SUSY breaking terms in the scalar potential are induced, 
 and the vacuum we discussed may be changed. 
Since our SUSY vacuum conditions are required at the GUT scale 
 much larger than the typical SUSY breaking scale, 
 the SUSY breaking effects are expected to be small. 
However, there is an interesting possibility 
 that this small effects may generate the $\mu$ term 
 for the doublet Higgs superfields. 
Indeed, analyzing the scalar potential 
 based on the minimal supergravity scenario, 
 we can find that non-zero $\mu$ term is really generated 
 through the VEV of $M$. 
Unfortunately, it is too small, $\mu \sim m_{3/2}^2/M_{GUT}$, 
 as expected in the dimensional analysis, 
 where $m_{3/2}$ is the electroweak scale gravitino mass. 
Therefore, we need another mechanism to provide 
 the sizable $\mu$ term. 
For this issue, the Guidice-Masiero mechanism \cite{GM} is remarkable, 
 since this mechanism automatically generates 
 the $\mu$ term and the $B$ parameter with the electroweak scale 
 through the SUSY breaking effects. 

Finally, although the doublet-triplet Higgs mass splitting 
 is simply realized through the strong gauge dynamics, 
 our model is somewhat complicated. 
It may be possible to construct a simpler dynamical model 
 with our mechanism. 
Also, in order to obtain a more complete model, 
 we have to construct a concrete SUSY breaking sector, 
 and combine it with the $SU(5)_{GUT}$ sector. 
These directions are worth investigating.

\acknowledgments

We would like to thank Yasuhiro Okada and Takeo Moroi 
for useful discussions. 
This work was supported by the JSPS Research Fellowships  
 for Young Scientists (R.K.). 


%
\end{document}